# The Vibe Shift in Software Engineering: Evaluating AI-Led Conversational Programming for Performance, Cognition, and Responsible Adoption

Sales G. Aribe Jr. [a,*], Louie Jay S. Labastida [a]

[a] *Information Technology Department, Bukidnon State University, Malaybalay City, Philippines*
*Corresponding author: *sg.aribe@buksu.edu.ph*

***Abstract*—This study evaluates Vibe Coding, an emerging AI-led conversational programming paradigm that enables developers to generate software through natural-language interaction with large language models (LLMs). Using a mixed-methods design, the study assessed performance efficiency, cognitive implications, and responsible adoption in comparison with traditional and AI-assisted coding environments. Thirty participants, including professional developers and advanced computing students, completed equivalent programming tasks under three experimental conditions. Quantitative data were analyzed using descriptive statistics and repeated-measures ANOVA, while qualitative data were examined through thematic analysis. Results show that vibe coding significantly improved development efficiency, reducing task completion time by 27% compared with traditional coding and 12% compared with AI-assisted coding. However, these gains were accompanied by lower maintainability indices and higher security vulnerabilities, indicating trade-offs in software quality. Usability results yielded a good rating (SUS = 71.4), while cognitive workload remained moderate (NASA-TLX = 55.5), reflecting reduced syntactic effort but increased linguistic reasoning. Thematic analysis identified trust calibration, loss of control, cognitive adaptation, and prompt-engineering strategy as key constructs. Notably, perceived loss of control was associated with increased security risks due to reduced transparency and validation of AI-generated outputs. Based on these findings, the study proposes a three-pillar framework for responsible adoption: (1) hybrid integration of human and AI capabilities, (2) human oversight and transparent accountability, and (3) context-aware deployment. Overall, vibe coding enhances productivity but requires critical oversight, reinforcing its role as a transformative yet transitional paradigm in software development.**

## I. INTRODUCTION

Software development has continuously evolved through paradigm shifts that redefine how humans conceptualize, design, and implement digital systems. From procedural and object-oriented programming to agile methodologies and, more recently, low-code and no-code platforms, each transformation has pursued greater abstraction, automation, and accessibility [1], [2]. The latest evolution is marked by the integration of artificial intelligence (AI) into the development workflow, transforming programming from purely syntactic manipulation to cognitive collaboration between humans and machines. Modern LLMs such as GPT-based systems now generate, debug, and optimize code through natural-language prompts, positioning AI as both co-developer and cognitive partner in software production [3].

Within this evolving landscape, Vibe Coding, also known as AI-led conversational programming, has emerged as a novel paradigm. Coined by Karpathy in 2025, vibe coding describes a behavior-driven, dialogue-based approach wherein developers express intent through natural language, allowing the AI to translate descriptions into executable code [4]. Instead of focusing on syntax, programmers iteratively "see stuff, say stuff, run stuff, and refine stuff" [5]. This conversational workflow promises to democratize programming by lowering barriers for non-technical users and accelerating prototype development [6]. However, despite its accessibility, the paradigm remains empirically under-examined, with limited studies assessing its performance, reliability, and cognitive implications compared to traditional or AI-assisted programming.

Unlike conventional AI-assisted coding tools such as GitHub Copilot, which operate primarily through inline code suggestions and predictive completion within an integrated development environment, Vibe Coding represents a fundamentally different interaction model. In AI-assisted coding, the developer retains syntactic control and incrementally accepts or rejects machine-generated suggestions. In contrast, Vibe Coding adopts a fully conversational paradigm, where developers express high-level intent through natural-language prompts, and the AI generates complete functional outputs based on semantic interpretation [7], [8]. This shift from token-level assistance to intent-driven generation repositions the developer's role from code executor to cognitive orchestrator, introducing new dependencies on linguistic precision, contextual framing, and interpretive validation [9], [10], [11].

Preliminary commentaries highlight both optimism and skepticism. Proponents emphasize its potential to enhance creativity and reduce entry barriers, while critics argue that AI-generated code introduces new risks in quality, security, and maintainability [12]. Moreover, cognitive studies suggest that conversational coding may reduce syntactic load but simultaneously increase prompt-engineering fatigue and dependency on AI interpretation [13]. These contrasting perspectives underscore a critical research gap: empirical evidence is needed to determine whether vibe coding constitutes a true paradigm shift or a transitional tool within the broader spectrum of AI-assisted development.

This ambiguity is further compounded by emerging challenges in AI-interaction literacy, where developers must acquire new competencies in prompt construction, semantic clarity, and critical evaluation of AI outputs [14], [15]. At the same time, recent studies have reported increasing concerns regarding the maintainability and security of AI-generated code, particularly when developers rely on opaque or poorly interpreted outputs [16], [17], [18]. These issues highlight the risk of "loss of control," where reduced transparency in code generation may lead to weaker validation practices and increased exposure to vulnerabilities. Addressing these challenges requires not only performance evaluation but also a deeper understanding of how cognitive and technical factors interact in conversational programming environments.

Responding to this gap, the present study empirically evaluates vibe coding as an AI-led conversational programming paradigm, focusing on its (1) performance efficiency and maintainability relative to traditional and AI-assisted approaches, (2) cognitive and trust implications for developers, and (3) guiding principles for its responsible adoption in software practice. Using a mixed-methods design, the research integrates quantitative performance metrics and qualitative insights to capture both technical effectiveness and human-AI interaction dynamics.

The significance of this study lies in its contribution to understanding how conversational AI reshapes the epistemology of programming, from code creation to cognitive collaboration. By proposing a three-pillar framework that emphasizes hybrid integration, human oversight, and context-aware deployment, the research aims to inform developers, educators, and policymakers about responsible strategies for incorporating AI in software workflows. Ultimately, this study positions vibe coding not as a fleeting buzzword but as an emerging frontier in human–AI partnership, inviting further discourse on the ethical, cognitive, and technical dimensions of the next era in software engineering.

### *A. Evolution of Programming Paradigms*

The evolution of programming paradigms has always reflected the tension between abstraction and control, between simplifying development processes and maintaining precision over computational behavior. Early procedural programming emphasized step-by-step logic and close alignment with hardware operations [19], [20]. The advent of object-oriented programming (OOP) redefined software as modular, reusable components, promoting maintainability and scalability [21]. Subsequent frameworks, such as agile development, shifted attention from rigid planning to iterative collaboration, enabling faster feedback cycles and user-centered design [22].

The recent surge of low-code and no-code platforms marked a democratizing turn, reducing the technical barrier for non-programmers to build applications through drag-and-drop interfaces [23]. Yet, these tools still rely on predefined templates, limiting flexibility. The next logical progression in this trajectory is the integration of AI into programming workflows, where systems can infer intent and generate executable code from natural language instructions [24]. This emerging mode of interaction transitions programming from a syntactic to a semantic process, focusing on outcomes rather than implementation details.

In this continuum, Vibe Coding represents a novel inflection point. It reframes programming as an interactive dialogue between human reasoning and machine inference, embodying what Samsyudin [25] termed the behavior-driven conversational paradigm. While it inherits the accessibility goals of low-code systems, it extends them through AI's cognitive capabilities, positioning the machine not only as a tool but also as a co-author in software creation.

### *B. Conversational AI and Code Generation Technologies*

The integration of LLMs such as OpenAI's Codex and GPT-based systems has accelerated the transition toward conversational programming. These models are trained on massive code repositories and natural language datasets, enabling them to generate code snippets, refactor functions, or explain algorithms in response to human prompts [26]. GitHub Copilot, for example, functions as an AI pair programmer that predicts developers' next lines of code, significantly reducing boilerplate.

Empirical evaluations have shown mixed results. Anand et al. [27] reported that code-generation tools improve productivity but sometimes produce syntactically correct yet semantically flawed code, requiring manual verification. Cotroneo et al. [28] warned that AI-generated outputs may contain security vulnerabilities and hidden biases inherited from training data. Similarly, Sarkar & Drosos [29] described vibe coding as "see stuff, say stuff, run stuff," emphasizing its intuitive workflow while acknowledging its susceptibility to misinterpretation.

Beyond efficiency, conversational programming introduces a shift in epistemology, from explicit problem-solving to interactive exploration [30]. Developers act less as code writers and more as prompt engineers, curating, validating, and steering AI behavior. Sapkota et al. [11] noted that the success of vibe coding depends on the developer's ability to formulate precise natural-language directives and interpret AI responses critically. This fusion of linguistic articulation and computational reasoning marks a departure from traditional programming literacy toward semantic literacy, where meaning construction replaces syntax memorization.

### C. Cognitive and Trust Implications in Human-AI Collaboration

The cognitive dimension of AI-assisted programming has become a crucial field of inquiry. While AI can automate repetitive tasks and reduce syntactic burden, it simultaneously introduces new forms of cognitive demand centered on semantic interpretation and prompt construction [31]. Recent studies indicate that although AI tools reduce perceived cognitive load, they may also promote cognitive offloading, potentially weakening critical thinking and deep engagement [32]. Rather than reiterating the syntactic-to-semantic shift, emerging research highlights the role of metacognitive engagement, where developers must anticipate AI behavior, evaluate outputs critically, and iteratively refine instructions through prompt engineering, now recognized as an essential skill for effective human–AI interaction [33]. This shift reflects a transition from procedural execution to cognitive orchestration, requiring higher-order reasoning, sustained attention, and adaptive thinking during human–AI collaboration [34].

Trust in AI systems further complicates this relationship. Only partial automation tends to foster optimal trust; over-reliance can lead to complacency, whereas under-trust results in under-utilization. In Drosos et al. [35] case study, participants appreciated conversational flexibility but still performed post-generation code reviews to ensure correctness. This aligns with Sison et al. [36] observation that working with LLM-based tools induces "intellectual vigilance," requiring continuous human oversight even as tasks become linguistically simpler.

From a psychological standpoint, vibe coding exemplifies a cognitive outsourcing phenomenon, where creative ideation and syntactic encoding are distributed between human and AI agents [37]. Such distributed cognition can enhance innovation but may also erode long-term skill retention if not balanced with deliberate practice. Therefore, evaluating vibe coding's cognitive implications extends beyond performance metrics to questions of human agency, skill sustainability, and epistemic dependence on AI systems.

### D. Responsible AI and Ethical Adoption in Software Development

As conversational programming systems gain prominence, responsible AI governance becomes essential. The European Union's (EU) AI Act and the Organization for Economic Co-operation and Development (OECD) AI Principles emphasize human oversight, transparency, and accountability, principles equally vital in software engineering contexts [38], [39]. The International Organization for Standardization and the International Electrotechnical Commission (ISO/IEC) 25010:2023, which defines quality characteristics such as reliability, security, and maintainability, provides a technical foundation for evaluating AI-generated software [40].

Beyond technical capability, the democratization of software development through AI must be understood within a broader ethical and governance context. While conversational programming lowers entry barriers and expands participation, it simultaneously introduces new responsibilities related to quality assurance, transparency, and accountability [41], [42].

Scholars advocate for hybrid development models that integrate human validation checkpoints into AI workflows [43]. These frameworks ensure that AI contributions remain auditable and that final accountability rests with human developers. Costa et al. [44] further argued that democratizing software creation must not compromise ethical standards; AI systems should empower creativity while preserving safety and inclusivity.

In educational and professional settings, responsible adoption entails teaching critical AI literacy [45], training developers not merely to use AI tools but to question, interpret, and refine their outputs. The adoption of vibe coding, therefore, must be contextualized within ethical boundaries that balance efficiency with reliability, and automation with human judgment. The three-pillar framework later proposed in this study, which is hybrid integration, human oversight, and context-aware deployment, emerges from this ethical imperative.

### E. Intellectual Property and Security Risks in AI-Generated Code

As AI-generated code becomes increasingly prevalent, concerns surrounding intellectual property (IP) and software security have emerged as critical dimensions of responsible adoption. Large language models are trained on vast corpora of publicly available code, which may include licensed or copyrighted materials, raising questions about ownership, attribution, and potential infringement when generated outputs resemble proprietary implementations [46], [47]. Recent legal scholarship highlights that generative AI challenges traditional copyright frameworks, particularly in determining authorship, originality, and whether outputs constitute derivative works [48]. Developers using conversational programming tools must therefore ensure compliance with open-source licensing obligations and institutional IP policies.

In addition to IP concerns, emerging studies identify specific security vulnerabilities associated with AI-generated code. These include insecure coding patterns, inadequate input validation, and susceptibility to prompt injection attacks, where adversarial prompts manipulate model behavior to produce unintended or harmful outputs [49], [50]. Empirical research further indicates that AI-generated code may replicate insecure practices embedded in training datasets, thereby increasing the likelihood of exploitable vulnerabilities if outputs are not rigorously reviewed and tested [51].

These challenges underscore the importance of integrating secure coding standards, automated vulnerability detection, and human oversight within AI-assisted development workflows. Without such safeguards, the efficiency gains of conversational programming may be offset by increased exposure to legal liabilities and cybersecurity risks. Consequently, responsible adoption must extend beyond performance metrics to incorporate legal, ethical, and security-aware development practices.

### *F. Synthesis and Research Gap*

The reviewed literature reveals a convergence of trends: (1) a historical move toward abstraction and accessibility in programming paradigms; (2) the rapid advancement of conversational AI tools for code generation; (3) the growing cognitive and trust challenges in human–AI collaboration; and (4) the urgent call for responsible governance of AI in software development. However, despite growing discourse on AI-assisted coding, empirical research that specifically evaluates "vibe coding" as a distinct paradigm remains scarce. Existing studies focus on usability or performance of tools like Copilot, but seldom examine the broader implications on cognition, maintainability, and ethical integration [52].

This gap highlights the necessity for a comprehensive evaluation that bridges technical performance and human factors. The present study addresses this need by employing a mixed-methods design to assess the efficiency, maintainability, cognitive workload, and user trust of vibe coding, culminating in a framework for responsible adoption. Through this approach, the study helps ground conversational programming not merely as a technological novelty but as a measurable and ethically accountable paradigm in modern software engineering.

## II. Materials and Methods

### *A. Research Design*

This study employed a mixed-methods research design to empirically evaluate Vibe Coding as an AI-led conversational programming paradigm. The design integrated both quantitative performance assessment and qualitative cognitive inquiry to capture the dual nature of software development, its measurable efficiency and its human-centered experience. A quasi-experimental approach was implemented to compare three distinct programming modalities: (1) traditional programming, in which participants created code manually without AI assistance; (2) AI-assisted programming, where participants used semi-automated coding tools such as GitHub Copilot; and (3) Vibe Coding simulation, which represented a fully conversational workflow conducted through natural-language interaction with an LLM.

This design followed Creswell and Plano Clark's [53] convergent parallel mixed-methods model, which allows quantitative and qualitative data to be collected concurrently, analyzed separately, and then merged for comprehensive interpretation. The goal was to determine not only how vibe coding performs technically but also how developers cognitively experience and interpret the paradigm. Through this combination, the study sought to assess the strengths and limitations of vibe coding as a practical and psychological innovation in software engineering.

### *B. Research Locale and Context*

The experiment was conducted within a controlled laboratory environment at the ACLC College of Bukidnon, Philippines, ensuring consistent access to development tools, internet connectivity, and standardized computing specifications. The study also simulated real-world software development conditions by using Integrated Development Environments (IDEs) such as Visual Studio Code to mirror the authentic experience of professional programmers. Ethical protocols were strictly followed in compliance with institutional research standards.

### *C. Participants*

The study involved thirty participants selected through purposive sampling to represent both industry and academic perspectives. The group comprised fifteen professional software developers with at least three years of programming experience and fifteen advanced computing students enrolled in an information technology program. This composition allowed the study to represent two perspectives: practitioners with established coding experience and learners familiar with emerging AI-assisted workflows.

All participants were proficient in Python and JavaScript, chosen for their widespread adoption in both academic and industrial AI-assisted programming environments. Informed consent was obtained from each participant, emphasizing data confidentiality, voluntary participation, and the right to withdraw at any stage.

### *D. Materials and Tools*

A consistent toolset was deployed to maintain comparability across all experimental conditions. The AI-assisted platforms included OpenAI's Codex-based systems (accessed through the ChatGPT interface) and GitHub Copilot, representing the baseline AI-assisted and conversational-AI environments, respectively. All development work was performed in Visual Studio Code, served as the unified IDE for all participants to maintain consistency and chosen for its extensibility and standardization.

Performance data were captured through automated logging scripts that recorded development time, number of errors, and debugging iterations. Code quality and maintainability were analyzed using SonarQube and CodeQL, which computed cyclomatic complexity, duplication metrics, and maintainability index (MI) scores, as well as detected security vulnerabilities.

For usability and cognitive assessment, the study employed the System Usability Scale (SUS) to measure perceived ease of use and satisfaction, and the NASA Task Load Index (NASA-TLX) to evaluate mental demand, effort, and frustration levels. Qualitative feedback was gathered through semi-structured interview guides focusing on trust, sense of control, and overall user experience when interacting with the AI systems.

### *E. Experimental Procedure*

Each participant was required to complete three programming tasks of equivalent complexity under the three conditions: traditional coding, AI-assisted coding, and vibe-coding simulation. The tasks were standardized to ensure comparability and included algorithmic logic, data handling, and user-interface routines. Specifically, the programming tasks consisted of three categories: (1) development of a command-line based inventory management system that supports item registration, updating, and search functionalities, (2) implementation of a data parsing and transformation module that processes structured input files (e.g., CSV/JSON) and generates formatted outputs with filtering and aggregation features, and (3) creation of an interactive form-based application with input validation and dynamic event handling. These tasks were selected to represent common real-world programming activities involving logic design, data manipulation, and user interaction. Task complexity was calibrated to ensure equivalent cognitive and technical demands across all experimental conditions.

Before beginning, participants attended an orientation session that explained the research objectives, ethical safeguards, and tool use. They also received short tutorials to guarantee equal familiarity with each platform. During the traditional coding condition, participants wrote source code entirely on their own, referencing only official documentation. In the AI-assisted condition, they relied on GitHub Copilot for suggestions while retaining full manual control of the code. In the vibe-coding simulation, participants interacted with the AI solely via natural-language prompts, describing desired behaviors and iteratively refining outcomes in response to feedback. To ensure consistency in the Vibe-Coding condition, participants were guided by a standardized prompt structure consisting of: (1) task description, (2) functional requirements, (3) constraints (e.g., programming language, coding standards), and (4) refinement instructions (e.g., validation, optimization, or modularization). A sample prompt used in the experiment is as follows:

*"Create a command-line-based inventory management system in Python that allows users to add, update, delete, and search for items. Ensure input validation, modular code structure, and error handling. Optimize the code for readability and maintainability."*

Participants were allowed to iteratively refine prompts in response to system responses, simulating real-world conversational programming workflows while maintaining experimental control.

Throughout all sessions, development time, error frequency, and debugging cycles were automatically logged. After completing each task, the code was tested for functional correctness using pre-defined test cases. Participants then completed the SUS and NASA-TLX questionnaires, followed by semi-structured interviews that explored perceptions of usability, trust, cognitive load, and control. Every session was video-recorded to document behavior, prompting strategies, and code-iteration patterns for subsequent analysis.

### *F. Data Collection and Metrics*

Data were collected across three major categories to ensure a comprehensive evaluation of the vibe-coding paradigm. The first category, performance metrics, included the total development time measured in minutes to determine how long each participant took to complete a task, the error frequency that recorded the number of syntax and logic errors encountered during coding, the number of debugging iterations completed before achieving a fully functional solution, the MI computed through static analysis tools to assess code quality, and the count of security vulnerabilities identified through automated scanning procedures.

The second category focused on usability and cognitive measures, which were quantified using validated instruments. The SUS provided numerical scores that reflected participants' perceived ease of use and overall satisfaction with each programming condition, while NASA-TLX captured multiple dimensions of cognitive workload, including mental demand, effort, and frustration.

Finally, the third category addressed qualitative perceptions, which explored the human factors associated with AI-led programming. Themes related to trust in AI, loss of control, cognitive fatigue, and prompt-engineering strategies were extracted from participant interviews and subsequently coded through systematic thematic analysis. Together, these data streams offered both objective and subjective insights into the performance, usability, and cognitive implications of vibe coding.

### *G. Data Analysis*

Quantitative data were processed using descriptive and inferential statistics. Mean values and standard deviations were computed for each metric, and a one-way repeated-measures ANOVA was conducted to test significant differences across the three programming conditions. Post hoc comparisons using the Bonferroni correction determined which pairs of conditions differed significantly, while effect sizes ($\eta^2$) were calculated to evaluate the magnitude of these differences.

Qualitative data from interviews were analyzed through thematic analysis following Braun and Clarke's [54] six-phase framework of familiarization, coding, theme generation, review, definition, and synthesis. Emerging themes such as trust in AI, cognitive effort, and perceived autonomy were identified and validated against quantitative results. The study employed a convergent parallel integration strategy, merging quantitative and qualitative findings during interpretation to uncover convergences and divergences between technical performance and cognitive perception.

### *H. Validity, Reliability, and Ethical Considerations*

Methodological rigor was ensured through multiple validation procedures. Internal validity was achieved by maintaining controlled experimental conditions that minimized extraneous variables. Construct validity was upheld using standardized instruments such as SUS and NASA-TLX, which accurately measured usability and cognitive load. Reliability was verified through pilot testing of all instruments, yielding a Cronbach's alpha of 0.87, which indicates high internal consistency.

Ethical considerations were strictly followed. Informed consent was obtained from all participants, and anonymity and data confidentiality were guaranteed. No sensitive or proprietary information was entered into AI systems during experiments. Ethical clearance was secured from the institutional review board, and the research adhered to principles of responsible AI use, including transparency, human oversight, and non-maleficence. All collected data were anonymized, securely stored, and used exclusively for academic purposes.

## III. Results and Discussion

### A. Performance Results

The first dimension of analysis focused on quantitative performance metrics from the three programming conditions: traditional coding, AI-assisted coding, and vibe-coding simulation. The evaluation examined five core indicators: development time, error frequency, debugging iterations, MI, and security vulnerabilities. Descriptive and inferential statistics, including a one-way repeated-measures ANOVA, were employed to identify significant differences among these modalities. A consolidated summary of the results is presented in Table 1.

TABLE I
SUMMARY OF PERFORMANCE EVALUATION ACROSS PROGRAMMING CONDITIONS

| Metric | Traditional Coding | AI-Assisted Coding (GitHub Copilot) | Vibe Coding Simulation | Statistical Outcome / Observation | Interpretation |
|---|---|---|---|---|---|
| Average Development Time | Baseline (100 %) | 12 % faster than traditional | **27 % faster than traditional** | $p < .05$ (ANOVA significant) | Conversational interaction reduces completion time by automating boilerplate code. |
| Error Frequency | Moderate; manual debugging required | Reduced by ≈ 15 % through AI suggestions | **Slightly higher than AI-assisted (≈ +8 %)** | Not significant ($p > .05$) | Speed improves but semantic errors emerge from ambiguous prompts. |
| Debugging Iterations | 4–6 cycles on average | 3–5 cycles | **2–3 cycles on average** | $p < .05$ | AI shortens debugging loops through auto-correction and contextual suggestions. |
| Maintainability Index | High (mean ≈ 75) | Moderate (mean ≈ 70) | **Lower (mean ≈ 60)** | $p < .05$ (significant difference) | Vibe-coded outputs show ≈ 15 % more duplication and 18 % greater complexity. |
| Cyclomatic Complexity | Low to moderate | Moderate | **High** | $p < .05$ | AI-generated code is denser and less readable, reducing maintainability. |
| Security Vulnerabilities | Minimal detected | Slight increase (+10 %) | **Highest (+22 %)** | $p < .05$ | Conversational AI introduces potential security risks requiring manual review. |
| Overall Performance Rank | 3rd (slowest but most secure) | 2nd (balanced) | **1st in efficiency, 3rd in security** | --- | Vibe coding offers speed and ease of use but sacrifices long-term quality and safety. |

The results indicate that development efficiency improved significantly under the vibe-coding condition. Participants completed their tasks an average of 27% faster than in traditional coding and 12% faster than in AI-assisted coding. The difference in completion time was statistically significant ($p < .05$), demonstrating that conversational interfaces substantially accelerate development by automating repetitive syntax and generating boilerplate components. Participants frequently remarked that they could “skip the tedious parts” and concentrate on logic verification. These observations support Vaithilingam et al. [3], who reported that LLM-based assistants enhance productivity by reducing syntactic friction.

Despite these efficiency gains, maintainability and security emerged as major trade-offs. The MI for vibe-generated code averaged 60, considerably below the AI-assisted (≈ 70) and traditional (≈ 75) scores and the difference was statistically significant ($p < .05$). Static analysis revealed greater duplication and higher cyclomatic complexity, suggesting that while AI-produced code is functional, it tends to be less modular and harder to interpret. Several participants noted that although the output “works,” it is “difficult to trace or refactor.” This aligns with Chong et al. [12], who found that AI-generated code often demands extensive post-generation refactoring to meet professional software-quality standards. An illustrative example of increased cyclomatic complexity in vibe-generated code is shown in Fig. 1.

```python
def process_data(data):
    if data:
        if isinstance(data, list):
            for item in data:
                if item > 0:
                    if item % 2 == 0:
                        print("Even positive")
                    else:
                        print("Odd positive")
                else:
                    print("Non-positive")
        else:
            print("Invalid data type")
    else:
        print("No data provided")
```

Fig. 1 Code snippet illustration on cyclomatic complexity

This structure demonstrates nested conditional logic that increases branching complexity, reducing readability and maintainability compared to modular or refactored alternatives. Such patterns were consistently identified in automated scans, supporting the observed decline in maintainability indices.

The security evaluation corroborated these concerns. Automated vulnerability scans detected approximately 22% more potential risks in vibe-coded outputs than in manually written programs. Most issues involved weak error handling, insecure dependency calls, or missing input validation routines. Specifically, the most frequently detected vulnerabilities included missing input validation (e.g., unsensitized user inputs), improper exception handling, and insecure use of external libraries without version constraints. For instance, several generated solutions accepted user input directly without validation checks, increasing susceptibility to injection-based attacks. Additionally, error-handling routines were often generalized (e.g., broad exception catching) without proper logging or recovery mechanisms, reducing traceability and robustness. These findings demonstrate that while conversational AI improves productivity, it can also introduce latent vulnerabilities if developers rely solely on AI outputs without rigorous verification. To provide a clearer understanding of the observed increase in vulnerabilities, Table 2 summarizes the most common issues identified through static analysis tools.

TABLE II
SECURITY VULNERABILITIES IN VIBE-GENERATED CODE

| Vulnerability Type | Description | Observed Impact |
|---|---|---|
| Missing Input Validation | Sensitive values embedded in source code | Increased risk of injection attacks |
| Hardcoded Credentials | Sensitive values embedded in source code | Exposure of security-critical information |
| Insecure Library Usage | Use of outdated or unverified dependencies | Potential exploitation of known vulnerabilities |
| Improper Exception Handling | Generic error catching without logging | Reduced traceability and debugging capability |
| Lack of Access Control | Missing role or permission checks | Unauthorized data access |

These findings indicate that while AI-generated code is functionally correct, it may omit essential security practices, reinforcing the need for manual validation and security auditing. In terms of error frequency and debugging iterations, vibe-coding sessions exhibited slightly more semantic errors than AI-assisted coding but still required the fewest debugging cycles overall, averaging only two to three iterations compared with four to six in traditional programming. This outcome suggests that while large-language models effectively prevent syntactic errors, they still require human intervention to resolve conceptual or contextual inaccuracies, particularly when prompts are ambiguous or underspecified.

Taken together, the quantitative results confirm that vibe coding excels in short-term efficiency yet compromises long-term maintainability and reliability. The paradigm demonstrates strong potential for rapid prototyping, instructional use, and exploratory development but remains unsuitable for mission-critical or security-sensitive applications without complementary human oversight. These results validate the first major claim of this study: AI-led conversational programming enhances productivity but must be integrated within a responsible, hybrid framework to preserve code quality, security, and sustainability.

## *B. Usability and Cognitive Performance Results*

The second dimension of analysis examined the usability and cognitive workload across the three programming conditions. Quantitative findings were derived from standardized instruments, namely the SUS and the NASA-TLX, while qualitative interview responses contextualized users' perceptions of ease of use, mental demand, and overall satisfaction. Together, these results illuminate how developers cognitively engage with vibe coding and how it compares to traditional and AI-assisted approaches.

TABLE III
SUMMARY OF SUS SCORES

| Programming Condition | Mean SUS Score | Qualitative Descriptor | Relative Rank | Interpretation |
|---|---|---|---|---|
| Traditional Coding | 65.8 | Marginally Acceptable | 3rd | Manual control ensures precision but increases procedural effort and reduces user comfort. |
| AI-Assisted Coding (GitHub Copilot) | 74.6 | Good Usability | 1st | Balanced automation enhances convenience while preserving user agency. |
| Vibe Coding Simulation | **71.4** | Good Usability / Engaging | 2nd | Natural-language interface improves intuitiveness and accessibility but depends on prompt clarity. |

As presented in Table 3 and Fig. 2, vibe coding achieved a mean SUS score of 71.4, placing it between traditional coding (65.8) and AI-assisted coding (74.6). Although slightly lower than AI-assisted performance, this rating still falls within the "good usability" range, indicating that participants found conversational interaction intuitive and engaging. Developers frequently described the interface as "easy to start with" and "less intimidating for non-experts." The qualitative interviews supported these findings, revealing that users appreciated the reduction in syntax-related stress and the immediate feedback loop generated through dialogue with the AI.

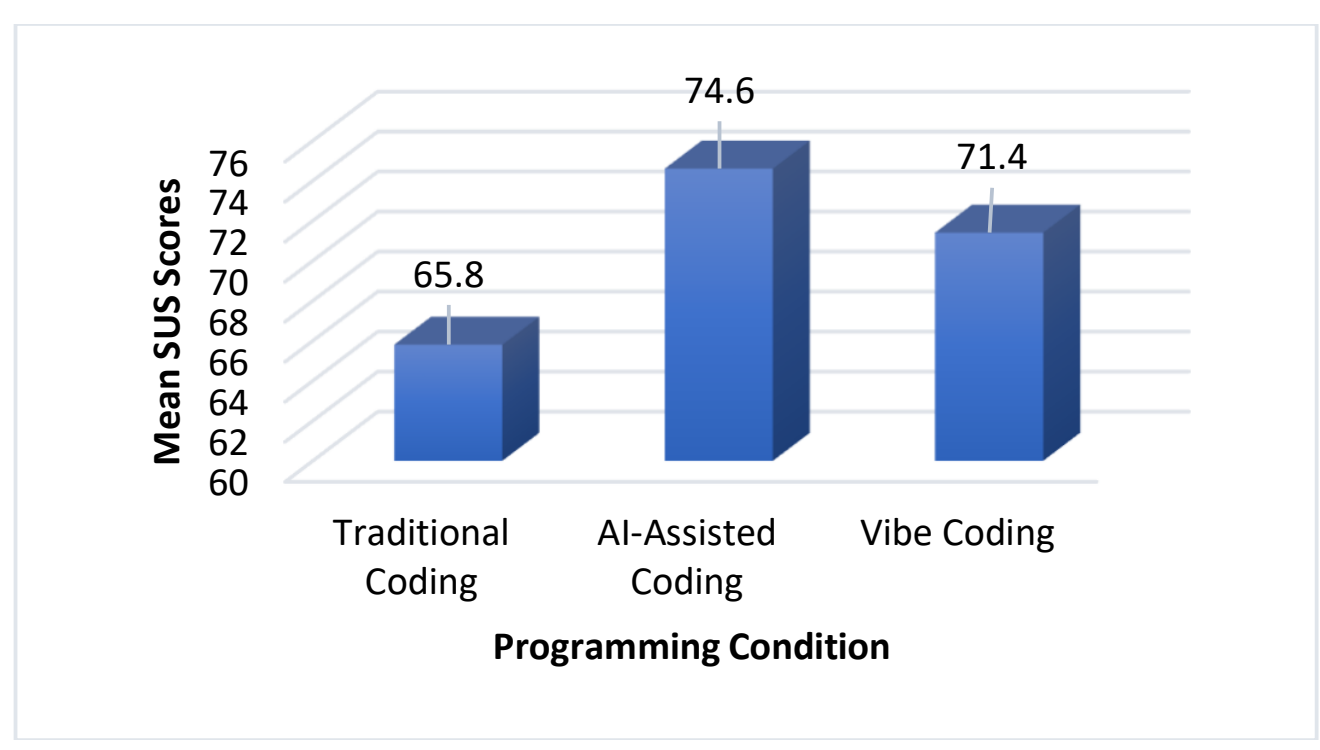

Fig. 2 SUS scores per programming condition

However, participants also reported occasional frustration when prompts were misinterpreted or when the AI produced syntactically correct but semantically irrelevant code. These inconsistencies lowered perceived predictability and trust, slightly affecting overall usability. This observation aligns with Hemdev [5], who noted that conversational programming, while intuitive, can challenge users to balance linguistic expression with computational precision. Overall, the results indicate that vibe coding's usability advantage lies in its accessibility and potential for engagement, particularly for novice developers and rapid-ideation contexts.

TABLE IV
SUMMARY OF NASA-TLX RESULTS

| Workload Dimension | Traditional Coding (Mean) | AI-Assisted Coding (Mean) | Vibe Coding Simulation (Mean) | Trend / Observation | Interpretation |
|---|---|---|---|---|---|
| Mental Demand | High (75 / 100) | Moderate (60 / 100) | **Moderate to High (65 / 100)** | Reduced syntactic recall but increased prompt formulation effort. | Cognitive effort shifts from coding syntax to crafting accurate natural-language prompts. |
| Physical Demand | Low (20) | Low (18) | **Low (17)** | Negligible differences. | All conditions involve minimal physical strain. |
| Temporal Demand | High (72) | Moderate (55) | **Lower (50)** | Time pressure decreases due to faster AI generation. | Conversational automation accelerates task completion. |
| Performance Satisfaction | Moderate (65) | High (80) | **High (78)** | Comparable to AI-assisted. | Participants were satisfied with the speed and convenience of vibe coding. |
| Effort Level | Very High (82) | Moderate (60) | **Moderate (63)** | Marked reduction compared to traditional. | AI eases repetitive workload, but demands focused communication. |
| Frustration Level | High (70) | Moderate (50) | **Moderate to High (58)** | Slightly higher in conversational mode. | Misinterpretation of prompts occasionally causes fatigue. |
| Overall Weighted NASA-TLX Score | ≈ 64.0 | ≈ 53.8 | **≈ 55.5** | --- | Vibe coding reduces workload compared with manual coding but remains cognitively demanding in prompt design. |

The NASA-TLX results (see Table 4 and Fig. 3) provide a nuanced picture of the cognitive dynamics involved in each programming condition. Vibe coding yielded an overall mean workload score of approximately 55.5, lower than traditional coding (64.0) but slightly higher than AI-assisted coding (53.8). This indicates that conversational programming effectively reduces mechanical and temporal load but introduces strategic mental demand due to the necessity of precise language articulation.

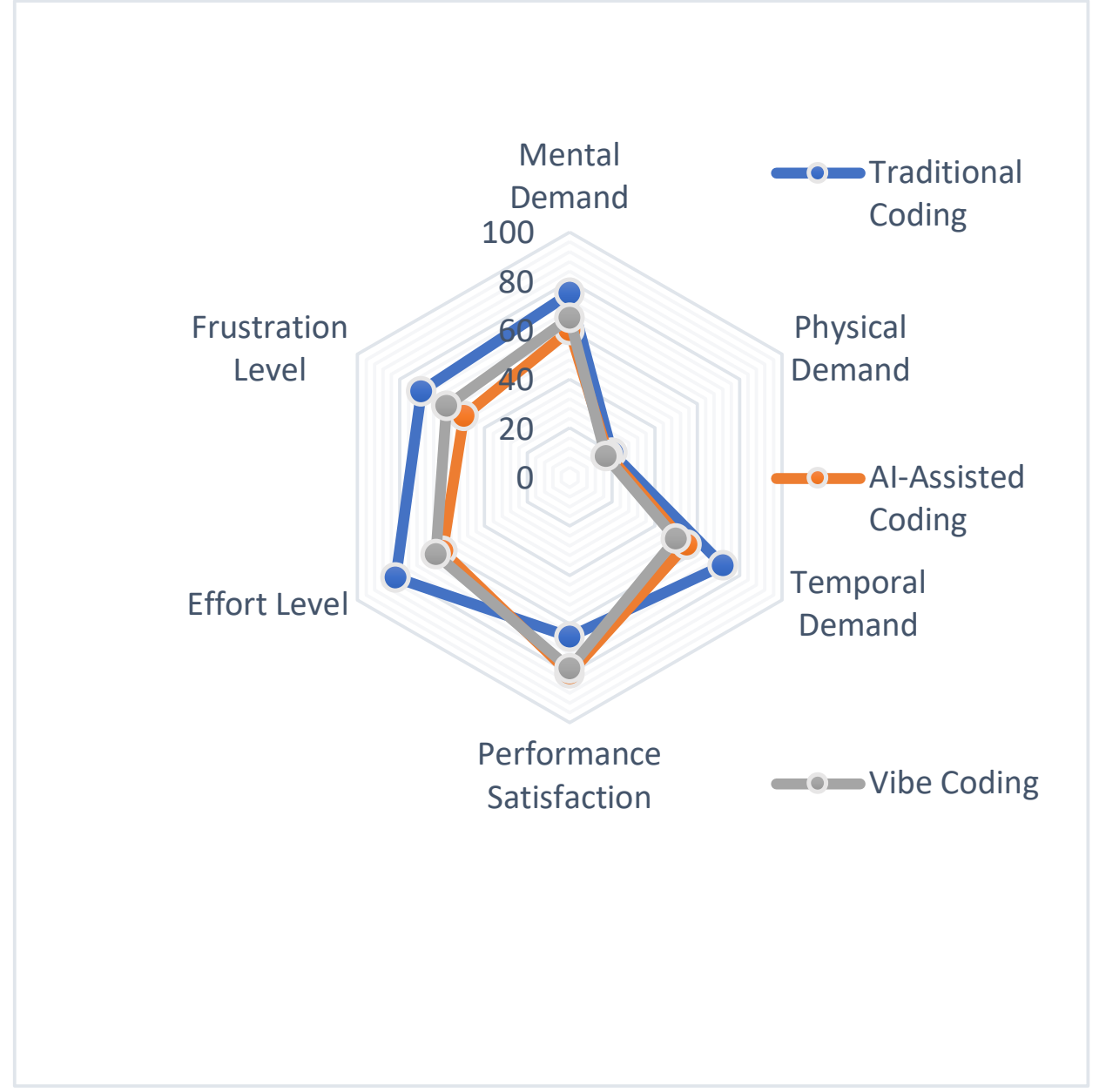

Fig. 3 Comparative NASA-TLX workload profiles across programming conditions

Interview data further clarified this pattern. Participants explained that while they no longer needed to memorize syntax or debug line-by-line, they had to "think like a teacher giving instructions to a student", anticipating how the AI might interpret their intent. Many referred to this as prompt fatigue, a cognitive strain arising from repetitive reformulation of inputs to achieve the desired result. Nonetheless, participants valued the AI's responsiveness, describing it as "creatively stimulating" and "collaborative." These results collectively suggest that vibe coding lowers the mechanical workload but raises the cognitive sophistication required for effective communication. The model relieves procedural burdens but demands metacognitive engagement, a shift from rote coding to interactive reasoning.

Overall, the usability and cognitive analyses confirm that vibe coding enhances user satisfaction and accessibility yet requires new forms of mental adaptability and linguistic precision. Developers must therefore cultivate AI-interaction literacy, an emerging competency involving prompt design, semantic clarity, and interpretive validation, to harness the paradigm's benefits without cognitive overload.

### C. *Cognitive, Trust and Security Implications*

The third analytical dimension of this study focused on the psychological and ethical facets of vibe coding, specifically, its effects on cognition, user trust, perceived control, and security awareness. While quantitative results revealed measurable performance advantages, qualitative feedback illuminated deeper patterns in how developers experience and manage collaboration with an intelligent coding partner. Thematic analysis of interview data generated four dominant themes: trust calibration, loss of control, cognitive adaptation, and prompt-engineering strategy. These themes are summarized in Table 5.

TABLE V
SUMMARY OF EMERGENT QUALITATIVE THEMES ON COGNITIVE, TRUST, AND SECURITY IMPLICATIONS

| Theme | Key Insights / Participant Expressions | Observed Implications | Interpretation |
|---|---|---|---|
| Trust Calibration | Developers expressed "partial trust" in AI outputs, relying on generated code for scaffolding but manually validating logic and security. | Selective adoption of AI suggestions; reliance on human verification before deployment. | Indicates dynamic trust management; users oscillate between dependence and skepticism. |
| Loss of Control | Participants felt "detached from authorship" when AI modified logic or produced unexplained solutions. Some described this as "AI hijacking my code." | Reduced sense of autonomy and accountability; emotional discomfort with opaque reasoning. | Highlights transparency as a prerequisite for sustainable adoption. |
| Cognitive Adaptation | Users reported "thinking in natural language" instead of algorithmic syntax; they described learning to "negotiate meaning" with the AI. | Shift from procedural to semantic cognition; expanded creative ideation. | Suggests reconfiguration of programming cognition toward linguistic reasoning. |
| Prompt Engineering Strategy | Participants refined prompting through iterative trial-and-error, embedding constraints and examples ("I had to specify what to avoid"). | Emergence of metacognitive skills: planning, prediction, and refinement. | Positions prompt engineering as a new literacy central to future programming practices. |

As shown in Table 4, trust calibration emerged as a defining cognitive and behavioral mechanism. Participants valued the AI's speed and convenience but consistently verified the generated code before integration, indicating a pragmatic, conditional trust rather than blind acceptance. Approximately two-thirds (67 %) of respondents stated that they used AI suggestions "only as a starting point." This behavior echoes Hemdev's [5] finding that developers engage in intellectual vigilance, a deliberate inspection of AI-produced code to preserve quality and accountability.

The theme of loss of control surfaced most strongly among experienced programmers accustomed to manual code manipulation. These participants reported unease when the AI autonomously altered syntax or introduced unfamiliar libraries without explicit instruction. Several described feeling "disconnected from authorship," while others worried about "not knowing why the code works." Such sentiments parallel the concern of Vandeputte [13], who warned that over-automation can erode developer agency and ownership. This perception underscores the ethical dimension of AI-driven collaboration: transparency, explainability, and auditability must accompany automation to sustain human oversight. It was particularly evident among professional developers, who expressed concern over the opacity and unpredictability of AI-generated outputs. Several participants described situations where the generated code diverged from their intended logic, requiring additional effort to trace and correct. As one participant noted, *"It feels like the AI is hijacking my code. I give instructions, but sometimes it takes over in ways I didn't expect"*. Another developer emphasized the verification burden, stating, *"I spend more time checking what the AI did than actually writing code myself, just to make sure nothing breaks"*. These perspectives reinforce the quantitative findings, which showed moderate cognitive load and reduced trust levels. The qualitative evidence indicates that while vibe coding accelerates development, it also necessitates continuous monitoring and validation, underscoring the importance of maintaining human oversight in conversational programming environments.

Cognitive adaptation represented the most constructive transformation observed. Developers began to shift their mental models from procedural syntax recall toward semantic intent formulation, effectively "thinking in natural language." This cognitive realignment fostered creativity and abstraction, enabling participants to conceptualize problems at a higher level. Yet this adaptation required significant effort; many described the early stages of vibe-coding interaction as "mentally taxing" until they internalized effective prompting techniques. The adaptation process, therefore, reflects both cognitive gain (through higher-order reasoning) and cognitive cost (through increased attentional demand).

The fourth theme, prompt-engineering strategy, encapsulates how participants transformed linguistic trial and error into a deliberate skill. They learned to embed contextual cues, specify constraints, and pre-empt ambiguities, behaviors that mirror problem-solving processes in human communication. This evolution signifies the emergence of a new digital literacy in programming: the ability to craft prompts that optimize AI comprehension and minimize semantic drift. As one participant summarized, "Prompting is the new debugging." Such skill development points to the hybridization of human-AI cognition, in which language replaces syntax as the core medium of control.

Finally, the interviews revealed cross-cutting concerns regarding security and data integrity. Participants questioned whether generated code might include insecure logic or undisclosed external dependencies. These apprehensions were validated by the quantitative results, which showed a 22 % higher incidence of vulnerabilities in vibe-coded outputs. Participants unanimously agreed that manual review and security scanning remain indispensable. This convergence between perception and evidence reinforces the necessity of human oversight in AI-mediated development.

In summary, the thematic evidence demonstrates that while vibe coding enhances cognitive flexibility and accelerates production, it simultaneously requires new forms of metacognitive awareness, ethical responsibility, and trust management. Developers must learn to collaborate with AI not as passive users but as critical partners, balancing creative autonomy with disciplined oversight. This cognitive and ethical recalibration defines the evolving landscape of human–AI co-programming, where responsible adoption depends on transparency, trust, and adaptive learning.

### D. Discussion

The integration of results from both quantitative and qualitative strands provides a comprehensive understanding of vibe coding as an AI-led conversational programming paradigm. Data drawn from performance metrics, usability and workload measures, and thematic insights reveal a consistent pattern: vibe coding significantly enhances development speed and user engagement but simultaneously introduces challenges in maintainability, cognitive management, and trust.

Quantitatively, the performance results demonstrated that vibe coding reduced development time by nearly 27% compared with traditional coding and 12% relative to AI-assisted programming. These improvements confirm its capability for rapid prototyping and time-efficient implementation. However, the corresponding decline in maintainability indices and elevated security-vulnerability rates underscore the trade-off between speed and software quality. The usability scores from the SUS likewise positioned vibe coding as accessible and engaging but slightly less predictable than AI-assisted environments, while the NASA-TLX indicated that conversational interfaces transfer effort from procedural syntax to linguistic reasoning. Together, these measures illustrate a hybrid pattern: mechanical effort decreases, yet cognitive complexity rises.

Another critical implication of reduced maintainability is the potential accumulation of technical debt. The lower maintainability index observed in vibe-generated code suggests that while initial development is accelerated, the resulting codebase may require additional effort for refactoring, debugging, and optimization in later stages. This introduces a temporal trade-off, where time saved during code generation may be offset by increased maintenance costs over the software lifecycle. Recent empirical studies indicate that AI-generated code frequently contains latent vulnerabilities and suboptimal structures that are not immediately detected during development, thereby increasing long-term maintenance burden and technical debt accumulation [55], [56], [57]

From a software engineering perspective, this indicates that vibe coding may shift effort rather than eliminate it, redistributing workload from initial development to post-development maintenance. Without proper governance, this could lead to compounding technical debt, particularly in large-scale or long-term projects where insecure or poorly structured code accumulates over time [58]. Therefore, integrating code review practices, refactoring cycles, and automated quality assurance mechanisms becomes essential to mitigate the long-term costs associated with AI-generated code.

The qualitative findings explain these numerical trends. Developers' accounts of trust calibration align with the empirical evidence of lower maintainability and higher vulnerability; users recognize that AI accelerates production but still require human validation before deployment. The feeling of loss of control corresponds with reduced trust and satisfaction scores, demonstrating that transparency in AI decision-making is critical for sustained adoption. Conversely, cognitive adaptation and prompt-engineering strategy help explain why participants achieved efficiency despite mental demand: once they internalized effective prompting techniques, interaction became faster and more intuitive. These insights reveal an emerging metacognitive equilibrium, in which users offset the AI's opacity with deliberate reasoning, linguistic precision, and oversight.

Integrating both strands confirms that vibe coding is a dual-edged innovation. Quantitatively, it represents measurable progress toward automated, user-friendly programming; qualitatively, it reshapes the developer's cognitive and ethical responsibilities. The intersection of these data sets supports three central interpretations: (1) efficiency without autonomy loss, (2) cognitive load evolves, and (3) trust. Efficiency without autonomy loss is conditional. Performance gains are attainable only when users maintain active verification and control over AI outputs. Cognitive load evolves, not disappears. Automation alleviates low-level tasks but introduces new layers of strategic thinking centered on prompt design and critical evaluation. Trust is an adaptive state, not a fixed attribute. Developers continuously negotiate confidence in the AI through iterative review, forming a pragmatic human-AI partnership rather than full delegation.

An important consideration emerging from this study is the long-term sustainability of developers' skills amid increasing reliance on conversational programming. As discussed in the literature review, AI-assisted development introduces a form of cognitive offloading that may affect skill retention and analytical depth [31], [32]. The present findings reinforce this concern. While vibe coding enhances productivity and supports rapid development, excessive dependence on AI-generated outputs may reduce opportunities for deliberate practice in core programming skills such as algorithm design, debugging, and code optimization.

Over time, this may diminish a developer's ability to critically evaluate or manually verify generated code, a capability that this study identifies as essential for maintaining software quality and security. This observation aligns with prior studies on human–AI collaboration, which emphasize the need for sustained human oversight and intellectual vigilance when interacting with AI-generated outputs [36], [59].

These findings suggest that AI-assisted programming should be implemented through a balanced, hybrid approach. Educational and professional practices must ensure that developers continue to engage in manual coding and validation tasks, thereby sustaining foundational competencies while leveraging the efficiency benefits of conversational AI. Such an approach safeguards long-term expertise while enabling responsible adoption of emerging programming paradigms.

In summary, the convergent analysis establishes that the success of vibe coding depends on a balanced synthesis of technical efficiency, cognitive adaptability, and ethical accountability. The paradigm cannot be evaluated solely by performance metrics; its true value lies in how effectively human judgment integrates with machine assistance. This holistic understanding provides the foundation for the proposed three-pillar framework for responsible adoption, which is discussed in the following section.

### *E. Proposed Framework for Responsible Adoption*

Drawing from the integrated findings presented in previous sections, this study proposes a three-pillar framework, as shown in Fig. 4, to guide the responsible and sustainable adoption of vibe coding as an AI-led conversational programming paradigm. The figure illustrates the dynamic and interdependent relationship among three core components: Hybrid Integration, Human Oversight, and Context-Aware Deployment.

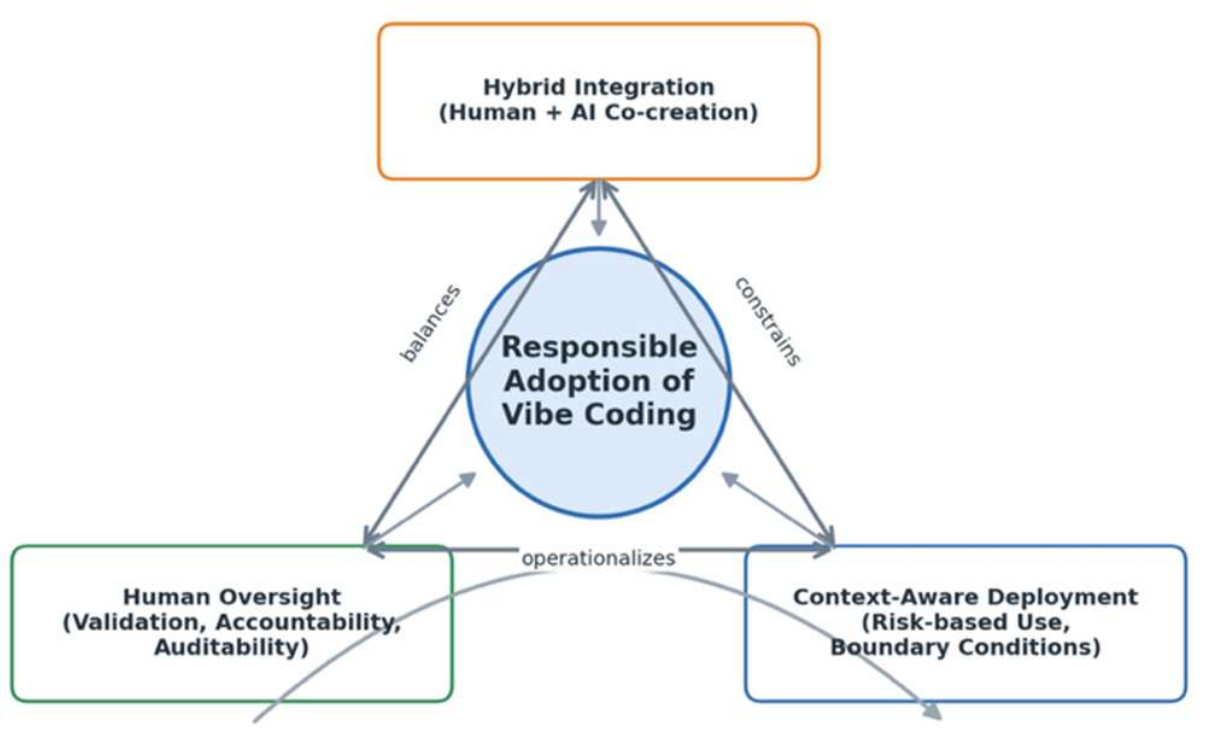

Fig. 4 Three-pillar framework for responsible adoption of vibe coding

At the center of the model is "Responsible Adoption of Vibe Coding," representing the desired outcome achieved through the coordinated interaction of the three pillars.

At the top, Hybrid Integration (Human + AI Co-creation) serves as the operational driver, enabling collaborative development between developers and AI systems. This pillar is connected to the other components through bidirectional relationships, indicating continuous interaction.

At the bottom left, Human Oversight (Validation, Accountability, Auditability) ensures that all AI-generated outputs are critically evaluated, validated, and aligned with ethical and quality standards. This pillar *balances* the system by maintaining human control over automated processes.

At the bottom right, Context-Aware Deployment (Risk-based Use, Boundary Conditions) specifies where and how vibe coding can be applied appropriately. It *constrains* the system by setting boundaries based on risk, application context, and criticality.

Human Oversight *operationalizes* Context-Aware Deployment. This means that validation, accountability, and auditability mechanisms are translated into enforceable deployment rules, boundary conditions, and risk-based usage decisions.

Arrows directed toward the central node indicate that each pillar contributes to achieving responsible adoption.

Finally, a curved feedback loop at the bottom illustrates the *continuous cycle of improvement across design, review, and deployment*, emphasizing that responsible adoption is an iterative and evolving process.

*1) Pillar 1: Hybrid Integration of Human and AI Capabilities*

The first pillar emphasizes the principle of hybrid integration, advocating for a balanced coexistence between AI-generated automation and human-directed validation. Findings from the performance evaluation show that vibe coding significantly accelerates task completion but reduces maintainability and increases vulnerability risk. To mitigate these trade-offs, AI assistance should be strategically confined to the early phases of development, such as rapid prototyping, ideation, and code scaffolding, where speed and experimentation are prioritized.

Human developers must remain central in later stages involving optimization, code review, and documentation. This hybrid strategy not only preserves software integrity but also aligns with ISO/IEC 25010:2023 standards on reliability and maintainability. In educational contexts, such integration can promote experiential learning, allowing students to witness the complementary strengths of human reasoning and AI-driven generation.

*2) Pillar 2: Human Oversight and Transparent Accountability*

The second pillar underscores the necessity of human oversight as a safeguard against algorithmic opacity and misplaced trust. Results from usability and trust analyses (Tables 2 and 4) revealed that developers maintained conditional confidence in AI outputs, often engaging in manual verification and correction. This behavior underscores the need for institutionalized review mechanisms, including peer evaluation, static code analysis, and automated security scans before deployment.

Human oversight ensures that AI-generated artifacts remain valid, auditable, explainable, and ethically accountable. Transparency must extend beyond functional correctness to include traceable decision logs and metadata showing how prompts led to specific outcomes. By institutionalizing these practices, organizations can reduce risk exposure while cultivating user trust, a critical factor for long-term adoption.

*3) Pillar 3: Context-Aware Deployment and Ethical Boundaries*

The third pillar advocates for context-aware deployment, recognizing that the suitability of vibe coding varies across domains. Empirical evidence demonstrated its strong performance in speed and accessibility but weaker reliability for mission-critical or security-sensitive applications. Consequently, adoption should follow a graduated model: exploratory use, assisted production, and restricted domain. Exploratory Use is for research, education, and creative prototyping. Assisted Production is under expert supervision and code-audit controls. Restricted Domains is subject to avoiding deployment in systems requiring certified safety, privacy compliance, or regulatory validation.

This stratified approach aligns with international AI-ethics frameworks, including the OECD AI Principles and the EU AI Act, both of which emphasize proportionality, human agency, and societal well-being. Ethical boundaries must also extend to data privacy and intellectual-property considerations, ensuring that AI training datasets and outputs comply with institutional and legal standards.

*4) Synthesis of the Framework*

The proposed framework operationalizes the study's central insight: that vibe coding's value lies not in full automation but in augmenting human creativity and judgment. It envisions a development ecosystem in which conversational AI serves as a cognitive collaborator rather than a replacement for human intellect. Hybrid integration preserves efficiency, human oversight guarantees accountability, and context-aware deployment secures ethical legitimacy. Together, these pillars form a responsible adoption model that balances innovation with control, reflecting the emerging paradigm of human-in-the-loop AI engineering.

## IV. Conclusions

This study evaluated Vibe Coding, an emerging AI-led conversational programming paradigm that enables developers to generate software through natural-language interaction with LLMs. Using a mixed-methods approach, the research examined its performance efficiency, usability, and cognitive implications, and ethical adoption framework relative to traditional and AI-assisted coding environments.

The results confirm that vibe coding represents both a technological innovation and a cognitive transformation in software development. Quantitatively, it delivered the highest efficiency, reducing average development time by up to 27% compared with traditional coding and 12% compared with AI-assisted coding. However, this gain came with trade-offs in maintainability and security, as code generated through conversational interaction exhibited higher duplication, greater complexity, and increased vulnerability rates.

Usability assessments revealed that vibe coding achieved a good usability rating (SUS = 71.4), indicating high potential for accessibility and engagement. Nevertheless, it also imposed a moderate cognitive workload (NASA-TLX = 55.5), as developers needed to articulate precise prompts and validate AI outputs. Qualitative findings enriched this analysis by identifying four core cognitive and behavioral themes, which are trust calibration, loss of control, cognitive adaptation, and prompt-engineering strategy, which describe how developers negotiate collaboration with an intelligent coding partner.

These findings underscore that vibe coding is a dual-edged paradigm: it democratizes programming and enhances productivity, yet it requires human oversight to maintain code integrity, transparency, and ethical accountability. The integration of quantitative and qualitative strands revealed that successful implementation depends on hybrid workflows in which human judgment complements machine automation.

To address these insights, the study proposed a three-pillar framework for responsible adoption encompassing (1) hybrid integration of human and AI capabilities, (2) human oversight and transparent accountability, and (3) context-aware deployment and ethical boundaries. This framework operationalizes the principle that AI should augment, not replace, human creativity and responsibility. It ensures that the benefits of conversational programming are realized without compromising quality, security, or ethical standards.

In conclusion, vibe coding signifies an evolutionary milestone in software development, a shift from code-centric to intent-driven creation. It highlights the growing need for AI-interaction literacy, enabling developers to craft precise prompts, critically assess AI behavior, and ensure responsible use of intelligent systems. While not yet a replacement for conventional programming, vibe coding serves as a powerful complementary paradigm that accelerates innovation while reinforcing the importance of human agency in the age of AI.

### A. *Recommendations*

Considering the findings and framework established in this study, several practical and forward-looking recommendations are proposed to guide developers, educators, researchers, and policymakers in adopting and advancing vibe coding responsibly. Developers and practitioners are encouraged to use vibe coding primarily as a supportive tool for rapid prototyping, concept testing, and documentation rather than as a stand-alone production platform. Continuous human oversight through peer review, automated testing, and code auditing should remain integral to mitigate the risks of logic inconsistencies and security vulnerabilities. As conversational interfaces redefine the developer's role, it becomes essential to cultivate AI-interaction literacy, the ability to formulate effective prompts, interpret AI outputs critically, and uphold ethical accountability in co-creation processes.

In the academic context, computing programs should integrate conversational AI programming modules into their curricula, emphasizing both technical skills and critical AI literacy. Educators should expose students to real-world applications of AI-assisted development while maintaining awareness of potential cognitive dependencies and intellectual-property considerations. Interdisciplinary collaboration among computer scientists, ethicists, and cognitive researchers is likewise recommended to better understand the psychological and pedagogical impacts of AI-assisted learning. For the research community, future investigations should focus on longitudinal effects of sustained exposure to conversational programming, particularly how it influences problem-solving ability, debugging skills, and cognitive dependence on AI. Expanding empirical studies to include team-based or enterprise-level settings could reveal how multiple human and AI agents interact in distributed development environments. Moreover, integrating explainable AI (XAI) frameworks into vibe-coding systems may enhance transparency, allowing developers to trace AI reasoning and decisions, thereby improving accountability and user trust.

Finally, policymakers and institutional leaders are encouraged to establish ethical and operational guidelines for AI-assisted software development that align with existing standards such as ISO/IEC 25010:2023 and the OECD AI Principles. Policies should promote transparency, data governance, and human accountability, ensuring that the deployment of vibe coding adheres to national and international ethical norms. Investment in capacity-building programs and research grants that support responsible AI innovation will further enable sustainable adoption.

In sum, the study recommends a balanced and forward-thinking integration of vibe coding into the software-development ecosystem, one that leverages the paradigm's strengths in creativity and efficiency while maintaining a strong foundation of human oversight, ethical awareness, and cognitive empowerment. Future work should continue to refine this equilibrium, ensuring that the evolution of conversational programming remains both technologically progressive and human-centered.

## Acknowledgment

We thank Bukidnon State University for the publication fee assistance of this research paper.